# Continuous agent-based modeling of adult-child pairs based on a pseudo-energy: relevance for public safety and egress efficiency

Chuan-Zhi (Thomas) Xie[abcd], Tie-Qiao Tang[a*], Alexandre Nicolas[b*]

a) School of Transportation Science and Engineering, Beijing Key Laboratory for Cooperative Vehicle Infrastructure Systems and Safety Control, Beihang University, Beijing 100191, China

b) Université Claude Bernard Lyon 1 and CNRS, Institut Lumière Matière, F-69622 Villeurbanne, France

c) School of Management and Economics, Beijing Institute of Technology, China

d) Yangtze Delta Region Academy of Beijing Institute of Technology, China

**Abstract**: Pushes, falls, stampedes, and crushes are safety hazards that emerge from the collective motion of crowds, but might be avoided by better design and guidance. While pedestrian dynamics are now getting better understood on the whole, complex heterogeneous flows involving e.g. adult-child pairs, though widely found at e.g. crowded Chinese training schools, still defy the current understanding and capabilities of crowd simulation models. We substantially extend a recent agent-based model in which each agent's choice of motion results from the minimization of a sum of intuitive contributions, in order to integrate adult-child pairs. This is achieved by adding a suitably defined pairing potential. The resulting model captures the relative positions of pair members in a quantitative fashion, as confirmed by small-scale controlled experiments, and also succeeds in describing collision avoidance between pairs. The model is used to simulate mixed adult-child flows at a T-junction and test the sensitivity to the design and pairing strategies. Simulation shows that making the post-confluence corridor wide enough is critical to avoid friction in the flow, and that tight hand-holding is advisable for safer evacuations (whereas more loosely bound pairs get split at high density) and, more marginally, more efficient egresses in normal conditions.

## 1. Introduction

Over the past two decades, crowd dynamics has been intensively studied, notably through the lens of public safety during evacuations, but

---

*Corresponding author:

tieqiaotang@buaa.edu.cn (T.-Q. Tang), alexandre.nicolas@cnrs.fr (A. Nicolas)

mostly for crowds made of single adults. However, according to the World Bank, the global population of children aged 0-14 years has been growing continuously from 1960 (1.13 billion) to 2022 (2.01 billion), and the top 5 countries in terms of total child population in 2022 are all developing countries [1]. Compared to developed countries, schools in developing countries are more likely to be overcrowded, according to an OECD report. For example, on average, in China, each class in primary school contains 37 children, while the corresponding figure for EU22 countries is only 20 [2]. In fact, multiple severe school stampede accidents have occurred in the after-class period in developing countries in recent years, e.g., (1) in 2017, a child died, and 21 children were injured in a primary school in Henan, China [3]; (2) in 2020, a stampede accident happened in a primary school in Kenya, which resulted in the tragic consequences of at least 14 students' deaths [4]. The above accidents show that the safety of moving crowds of students during the after-class period in various types of schools (e.g., kindergartens, primary schools) in developing countries should be better safeguarded. In particular, extracurricular training schools are ubiquitous in China; yet, according to one survey conducted by the Chinese Ministry of Education in 2018, around 8% of the existing training schools raise major safety hazards [5]. In such schools, pairs of adults and children are ubiquitous [6] (see Fig. 1(a)), which leads to significant differences in the global crowd dynamics compared to the more homogeneous crowds encountered in ordinary primary schools. In other words, the mixed crowd movement safety in training schools must be studied independently of ordinary primary schools.

Currently, in their pragmatic endeavor to curb crowd-movement safety hazards (e.g., stampedes), researchers have mostly resorted to two approaches: post-accident investigation and pre-accident prevention. Post-accident investigation relies on the analysis of multi-source data of stampedes, including videos and medical records. For instance, Helbing et al. [7] analyzed the video record of crowd crushes in Mina/Makkah during the Hajj in 2006. These studies may help detect the onset of an accident before it becomes tragic and provide a quick response to situations of stampede provided that they are similar enough to those already investigated. However, while crowd crushes do occur in school environments, the visual documentation and detailed record-keeping of these accidents appear relatively sparse compared to more publicized events, in particular those during the Hajj [7, 8] or the recent Halloween crowd crush in Seoul [9]. This scarcity in visual evidence may be ascribed to the complexities inherent in capturing information in such scenarios. In this regard, researchers are led to put more emphasis on the early prevention of accidents, notably by i) modeling pedestrian's microscopic behavior in typical settings inside schools (e.g., classrooms [10], staircases [11]); ii) identifying factors that bring hazards to crowd movement based on models' simulating results. For example, putting the focus on children's movement on various considered setups (e.g., corridors, bottlenecks, classrooms), researchers conducted empirical studies aiming to provide reliable data sources in order to bring forth, validate, and evaluate pedestrian flow models by means of questionnaire surveys, evacuation drills, controlled experiments, etc. [12-17]. Furthermore, since microscopic models [18-20] can integrate the characteristics of children's movement, simulating them can unveil potential safety risks, especially in scenarios which are hardly amenable to empirical studies. Naturally, when trying to improve crowd safety during the after-class period in training schools from both the perspective of pedestrian movement guidance and the architectural design, the accurate modeling of the adult-child mixed flow will be a prerequisite.

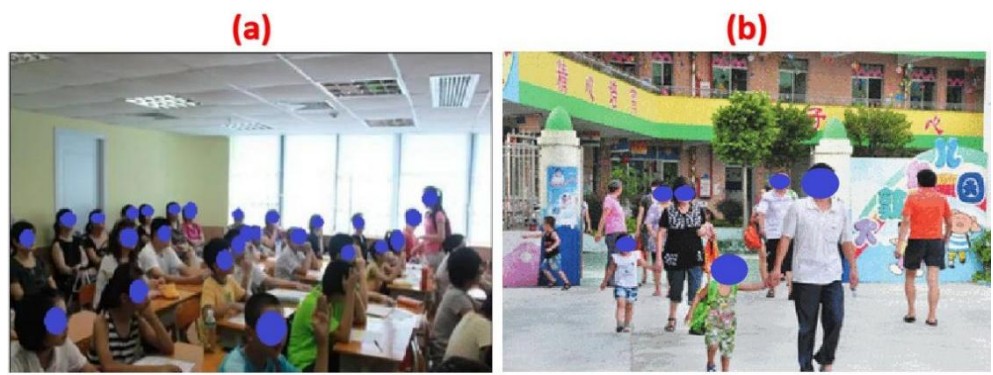

Fig. 1 (a) Photograph taken in a training school classroom, evidencing the heterogeneous crowd composition. (b) Snapshot of adult-child pairs holding hands.

Concerning the adult-child mixed pedestrian flow modeling, outdoor spaces (e.g., crossings and roads near schools) [21, 22] have received more attention than indoor facilities. Although adult-child pairs' microscopic behavior in outdoor and indoor spaces has certain similarities (see for instance the matching behavior depicted by Li et al. [23]), such pairs' behavior in a broader range of indoor scenarios (including training schools) undoubtedly deserves further exploration. Previously, we have discussed the adult-child pairs' movement in several scenarios present in training schools, including staircases [24] and classrooms [6], but not in unobstructed corridors. In the latter scenario, adult-child mixed crowds in corridors are less influenced by the built environment, but interactions between pedestrians are substantial. The typical hand-holding behavior (see e.g. Fig. 1(b)) results in a tight, but flexible 'bond' between pair members; this is expected to impact the collision avoidance maneuvers and more generally the flow in the bi-directional or confluence settings that are often observed in the after-class period. Existing models are not designed to reproduce these features and thus fall short of describing this complex situation. This paper intends to fill these gaps by proposing a versatile 'pseudo-energy' agent-based model that manages to describe the adult-child mixed flow in training schools with diversely shaped corridors.

The paper is organized as follows. Section 2 provides a thorough description of the spatially continuous 'pseudo-energy' based model with the integration of adult-child pairing characteristics. Then, Section 3 introduces key parameters of adult-child pairs walking (i.e., desired pairing distance, etc.), which are extracted from a controlled experiment, and exposes the simulation performances of the model. Finally, in Section 4, some corridor-design and crowd-guidance strategies are implemented and tested numerically.

## 2. Methods: description of the model

In this section, we present a continuous agent-based model for pedestrian dynamics which is robust (in terms of reliability of the predictions) and flexible (in terms of the range of scenarios and crowds that can be handled), but remains conceptually intuitive. In essence, the model interprets commonly observed phenomena in adult-child flows, such as pairing, collision-avoidance, as the results of the minimization of an abstract generalized cost, or 'pseudo-energy', made of different contributions. The decision-making process of each agent is thus rendered by an optimization process, rather than an equation of motion based on Newton's law or a variant thereof. Successfully introducing pairs (or, more generally, groups) in this framework is a major contribution of the present manuscript.

2.1 Review of existing models

To complete our overview of existing modeling frameworks, we searched for the following three keywords: child, pairing behavior, and agents' anticipation, as they are intrinsically aligned with our goals. In the search for pedestrian flow models involving children, we found that discrete models (e.g., Cellular Automata [25]), semi-continuous models (e.g., Optimal Step Model [26]), and continuous models (e.g., Social Force Model [27]) were widely adopted. Chen et al. proposed an extended CA to model children's non-emergency classroom evacuation process [28], and the model results closely matched their previous experimental data in a fairly simple classroom scenario structured by desks and chairs; based on the Optimal Step Model, Xie et al. accurately characterized the movement of adult-child mixed flow on stairs, and further tested flow-improvement strategies using an calibrated and validated EOSM [24]; generic commercial software based on continuous models, e.g., *Pathfinder*, is also commonly used to delve deeper into children evacuation/drill data [29, 30]; for instance, Cuesta et al. compared the simulation output results of software with empirical data, and found that in terms of both the ability of independent users to configure the models and the accuracy of the models examined, *Pathfinder* performed encouragingly [29]. From the above models, the modeling of crowd movement in different scenarios naturally calls for approaches with varying degrees of continuity; obviously, given the configuration features of the considered scenario (unobstructed corridors) and the corresponding walking flexibility within such scenarios, the continuous approach is more suitable.

An additional source of complexity of the adult-child mixed pedestrian flow is the pairing behavior of one adult and his/her child, which can be easily identified in daily life owing to the adult-child hand-holding. One of us comprehensively summarized the models accounting for ordinary pedestrian groups [31], but the hand-holding behavior makes the native version of models introduced by Ref. [31] unsuitable for adult-child pairs. Although adults often lead their children when walking, physical hand-holding contact resulting in the leading way can hardly be classified into the 'leader-follower' pattern introduced by Xie et al. [32, 33], where the structures of dyads are invisible. Our previous endeavors partly reveal how adult-child pairs walk inside classrooms and on stairs by empirical and modeling methods (i.e., through extending CA and OSM) [6, 24]; various hand-holding behaviors are both discovered in observations and reproduced by models in Refs. [6] and [24]. Yet, as the continuous approach has been selected to model the mixed flow in unobstructed corridors, Refs. [6] and [24] cannot answer how to incorporate the adult-child pairing effect into continuous pedestrian flow models. In terms of this, under a dilute environment, inspiration could be drawn from Zanlungo et al. since their work systematically provides a potential-based modeling method for depicting ordinary dyads and triads [34]; at relatively high-density conditions, Han and Liu integrated information transmission mechanism into the basic social force model for simulating group walking behavior [35]. Still, models in Refs. [34] and [35] are oblivious to the hand-holding feature commonly observed in adult-child pairs; thus, in this paper, 'pseudo-energy' based continuous model is brought forward.

In the after-class period of training schools, adult-child pairs would inevitably face bi-directional or confluent interaction with others; under these conditions, for ordinary pedestrians, both controlled experiments and observations have revealed the collision-avoidance behavior [36, 37]. By comparing the descriptive potential of two distinct variables, namely the ***Euclidean Distance*** and the ***Time to Collision*** (TTC), in real-life video recordings, Karamouzas et al. found that the TTC better renders the collision-free intention of pedestrians than *Euclidean Distance*, with notably a lower sensitivity to the rate of approach: pedestrians approaching each other slowly (rate-of-approach $\leq 1$-m/s) tend to stand closer on average than those with medium ($(1,2]$-m/s), or high ($\geq 2$-m/s) rate-of-approach; in contrast, through the lens of the TTC, a more homogeneous picture emerges for different rate-of-approach (see Fig. 1 in Ref. [37]). Karamouzas et al. also assumed, measured, and validated one 'virtual' energy (i.e., interaction energy), determined by TTC between agents and governing pedestrians' avoidance behavior [37]. Turning to the modeling of pedestrians' anticipation/avoidance, one can easily find that both the velocity-based model [38-40] and force-based model [41] have been adopted; for example, Xu et al. successfully reproduced crowd movement in both bottleneck and bi-directional corridor, through the validation of collectively indicators (e.g., fundamental diagram) [38-39]. Although existing models, including Ref. [39-41], are able to capture the collision-avoidance among adult-child pairs to some extent, we turn to the 'pseudo-energy' perspective in this work for the following substantiated reasons: i) a straightforward, but quantitatively realistic description of collision avoidance is possible using the TTC-based collision energy derived by Karamouzas et al., which proved its validity in a wide range of situations, speeds, and densities [37]; ii) the characteristics of pair-motion, including the fluctuations away from the mean pairing distance, can also be satisfactorily captured by a (pairing) pseudo-energy [34]. For collision avoidance between a dyad and an individual, more specifically, the recent work of Gregorj et al. based on two empirical datasets provides illuminating insight and argues in favor of a physics-inspired continuous potential energy to describe the relative distances at avoidance [43]. This work confirms the dependence of the avoidance behavior on the social relationship between pair members, with a lower probability of intrusion into the group as the interaction intensity between its members is stronger. Interestingly, the dyads of family members (mostly parent-child pairs) found in this Japanese dataset belong to the category with weaker interactions, hence less bonding;

however, it has been reported that these pairs often have a fluctuating spatial structure, which is quite different from the more strongly bonded adult-child pairs typically observed in Chinese training schools.

2.2 Introduction of a 'pseudo-energy' based model

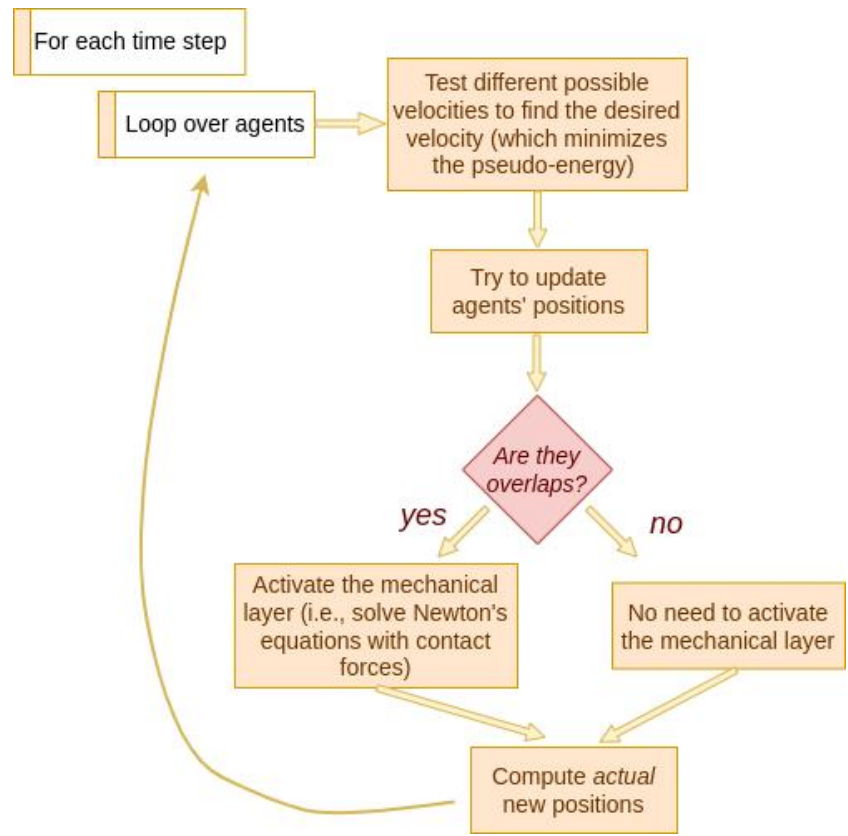

Fig. 2 Functional diagram ('workflow') of the 'pseudo-energy' based model.

At the heart of the 'pseudo-energy' based model is the decomposition of the pedestrian's choice of motion into successive stages of perception, decision, and execution. Here, we assume that decision is based on six factors that can be gauged by perception or proprioception, and which we transcribe quantitatively into pseudo-energies: i) a biomechanical cost associated with the test velocity; (2) spatial preferences (irrespective of the rest of the crowd) expressed as a static floor field; (3) an inertial cost; (4) a 'proxemics' cost associated to the violation of the private space; (5) a penalty for anticipated collisions; (6) a pairing pseudo-energy. Decision-making is then regarded as an optimisation process for the sum of these contributions, that is, pedestrians are supposed to aim for a next position associated with the minimum overall 'pseudo-energy'. Eventually, execution will be based on the resulting desired velocity. This however raises the question of how to proceed in the event of a collision. Should agents come into contact, the decision layer will not suffice and the mechanical layer is designed to handle the bouncing process between pedestrians who have already collided. In light of the above introduction, the 'workflow' of the proposed model accounting for the adult-child mixed pedestrian flow is shown in Fig. 2.

For individual pedestrians, the first five energies listed above, together with the mechanical layer algorithm, form the basis of microscopic modeling framework for pedestrian flow recently introduced in Ref. [44] and validated for a wide range of scenarios. Extending it to adult-child mixed pedestrian flows with a pairing energy is critical to our endeavor to address safety hazards at training schools. Let us now turn to a more technical description of the model components.

2.2.1 Decisional layer

The decisional layer determines, for each agent $i$, a desired velocity $u_i^*$. This velocity is regarded as the one which optimizes a perceived cost, or 'pseudo-energy', $E[u_i]$. One is thus left with the task of specifying this pseudo-energy. Since the movement choices of a pedestrian result from distinct needs and constraints, the pseudo-energy will comprise several contributions, which are here summed linearly. Let us briefly recall the main features of this function; for a detailed presentation of the model for single pedestrians, the reader is referred to Ref. [44].

$$E[u] = E_{biomech} + E_{FF} + E_{inertia} + E_{private-space} + E_{anticipation}. \quad (1)$$

In free space, only the biomechanical contribution $E_{biomech}$, measuring the (empirical) physiological cost of walking at a given speed $u = \|u\|$ [45], the static floor field $E_{FF}$ [1] (evaluated at the position $r' = r + \delta t \cdot u$ that would be reached at the next time step with the test velocity $u$), and the quadratic penalty $E_{inertia}$ for changing velocities too abruptly are active.

---
[1] The vicinity to walls is penalized in the evaluation of the static floor field $E_{FF}$. The characteristic repulsive length ($d_c$) is set as 0.15-m in the following, which slightly differs from the value chosen in Ref. [44].

Static floor field: rather than prescribing a global target or a series of way-points, the model is thus based on a floor field $E_{FF}[r]$ which gives, for any position $r$, how advantageous this location is for the pedestrian. Technically, this field is computed by solving an Eikonal equation with an A* algorithm; the Eikonal equation gives the optical path of a ray of light traveling through media of possibly varying refractive indices, which here denote the comfort of a given location (e.g., it is penalized very close to a wall).

To get clearer insight into the effect of the combined contributions of Eq. (1), consider the simple case of uniform motion of an isolated pedestrian. Then, only the contributions of $E_{biomech}$ and $E_{FF}$ are nonzero and the chosen velocity is directed along the gradient of $E_{FF}$ and its magnitude $v^{pref}$ minimizes the sum of the first two contributions. Accordingly, if one knows an agent's free-walking speed $v^{pref}$, the slope of the floor field can directly be obtained and the model contains no adjustable parameter at this point.

On top of these three contributions, for pedestrians walking **alone** (no pairs), interactions with the built environment and the crowd generate two new terms, reflecting two distinct types of repulsive interactions at play in pedestrian dynamics. The first one, $E_{private-space}$, is based on the separation distance between an agent and their neighbors, with a short-ranged repulsive strength decaying with distance, which is familiar to physicists. More precisely, we set:

$$E_{private-space}[r] = \sum \frac{\eta}{\sigma_i + \sigma_j} \cdot V^{rep}\left(\frac{\|r - r_j(t + \delta t)\|}{\sigma_i + \sigma_j}\right), \qquad (2)$$

where $\eta$ is a repulsion strength coefficient; $\sigma_i$ and $\sigma_j$ denote body radii. $V^{rep}(r)$ is set as:

$$V^{\text{rep}}(r) = \begin{cases} \dfrac{1}{r} - \dfrac{1}{1+\varepsilon^*} & \text{if } r < 1+\varepsilon^* \\ 0 & \text{otherwise} \end{cases}, \tag{3}$$

and it reflects the desire of people to preserve a private space around themselves, whose extent $\varepsilon^*$ may vary between individuals and between cultures (as studied by the field of proxemics). In the following Chinese adult-child mixed flow research case, spatial extent of the private space (relative to body width), $\varepsilon^*$, is set as 0.5. Beyond these concerns for private space, pedestrians also pay particular attention to the risk of future collisions and adapt their trajectories to avoid them. By using empirical data sets, Karamouzas et al. demonstrated that these effects are much more readily described using a new variable, the anticipated ***Time to Collision*** (TTC), than distances [37].

The TTC-based energy introduced in our model, $E_{\text{anticipation}}$, follows the same expression as that defined by Karamouzas et al. except that non-mechanical collisions between private spaces are also taken into account (which results in a smoother profile) and only the most imminent collision is considered.

This search for an optimal velocity $u_i^*$ minimizing $E[u_i]$ is performed every $\delta t$ seconds (set as 0.1-s in this paper), which corresponds to the aforementioned ***cognitive reaction*** time. In particular, it may happen that the selected desired velocity $u_i^*$ leads to a collision within $\delta t$ and thus activates repulsive mechanical forces; then the actual velocity can be computed using mechanical equations of motion.

2.2.2 Mechanical layer

Despite the wish of pedestrians to keep a safe distance from each other, contacts may arise in dense or competitive settings and should thus be suitably modeled. This is the aim of the mechanical layer, which treats pedestrians as self-propelled disks, with a propulsion velocity $u_i^*$ given by the decisional layer, and interacting with frictionless Hertzian interactions with 0.2-s as the relaxation time ($\tau^{\text{mech}}$). In Newton's equations, this yields:

$$m\ddot{\boldsymbol{r}}_i = m(\boldsymbol{u}_i^* - \dot{\boldsymbol{r}}_i)/\tau^{mech} + \sum_j \boldsymbol{F}_{j \to i}^{Hertz} + \sum_{w \in walls} \boldsymbol{F}_{w \to i}^{Hertz}, \quad (4)$$

where $\sum \boldsymbol{F}_{j \to i}^{Hertz} = -dU(r_{ij})/dr_{ij}$ with $U(r) = A \cdot \max(0, \sigma_i + \sigma_j - r)$ is a Hertzian contact force[2]. One can refer to Ref. [44] for a more specific explanation of the mechanical layer.

2.2.3 Integration of the adult-child pairing effect

Now, to be relevant in the practical situation of interest, the foregoing pseudo-energy defined by Eq. (1) must be complemented by a pairing energy $E_{pairing}$, viz.,

$$E[\boldsymbol{u}] = E_{biomech} + E_{FF} + E_{inertia} + E_{private-space} + E_{anticipation} + E_{pairing}. \quad (5)$$

We should note that within each pair, only $E_{pairing}$ will govern the interactions between adult and child, excluding the contributions $E_{private-space}$ and $E_{anticipation}$, which are set to zero between pair members. To define the pairing energy, we draw much of our inspiration from the interaction force specified by Zanlungo et al. [34], which is made of two key physical factors: the inter-agent distance and angle within one pair. We adopt the same factors here, but handle the problem in terms of energies rather than forces. The agents' velocities $\boldsymbol{u}_i$ and $\boldsymbol{u}_j$, the inter-agent distance $d_{i \to j}$ and the angle $\alpha_{i \to j}$ are defined in Fig. 3(a) for a pair, consisting of agent *i* (the adult) and agent *j* (the child), from the adult's perspective of agent *i* at the current time $t$. As a matter of fact, in the model, agents make their decisions based on the configuration that they expect at a subsequent time $t + \delta t$. Of course, their expectations can only be based on their current observations; therefore, the future positions of the other agents (say, of *j*, from *i*'s perspective) are computed using their current velocities $\boldsymbol{u}_j(t)$. In addition, hand-holding sets an upper bound on the possible pairing range: the maximum inter-agent distance (shown as a black line in Fig. 3(b)) cannot exceed the sum $d_{\max}(i,j)$ of arm lengths of agent *i* and agent *j*. Furthermore, the inter-agent angle is constrained by the chosen side for hand-holding, as soon as the hand-holding pair has formed; thus, $\alpha_{i \to j}$ should be either always positive or always negative. Combining the above two restrictions, Fig. 3(b) highlights the half-disk corresponding to the possible pairing range for the configuration of Fig. 3(a).

---

[2] *A* denotes agent's surface area. Eq. (4) is solved by a velocity Verlet algorithm.

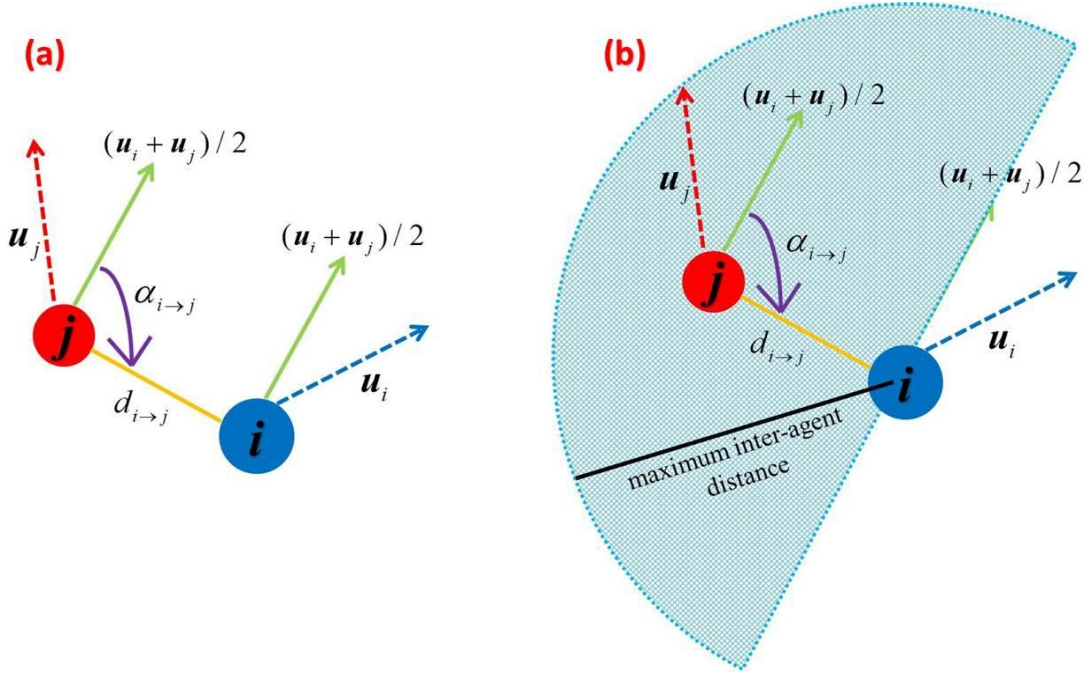

Fig. 3 (a) Geometric sketches and notations for an adult (*i*)-child (*j*) pair; (b) sketch illustrating the possible pairing range based on (a). The blue and red dashed arrows denote the velocities of agent *i* and agent *j* with symbols of $u_i$ and $u_j$; the green arrow shows the mutual walking direction for the pair, which is calculated by $(u_i + u_j)/2$; $d_{i \to j}$ (the yellow line) denotes the inter-agent distance. $\alpha_{i \to j} \in (-\pi, \pi]$ (the purple arrow) indicates the inter-agent angle, which is directly related to the group's walking direction and not to the individual ones (i.e., only *i*'s or *j*'s).

Using the notations of Fig. 3, the basic formula of $E_{pairing}$ is set as:

$$E_{pairing} = \begin{cases} c_d \cdot \left( \dfrac{d_{i \to j}}{d_0} + \dfrac{d_0}{d_{i \to j}} \right) + c_\alpha \cdot \left( (1+\omega)\alpha_{i \to j}^2 + (1-\omega)(\alpha_{i \to j} \pm \pi)^2 \right) & \text{if } d_{i \to j} \leq d_{\max}(i,j) \\ +\infty & \text{otherwise} \end{cases}, \quad (6)$$

where $d_0$ is the desired pairing distance of the pair and $c_d = 2$, $c_\alpha = 1$ are coefficients weighting the inter-agent distance and angle, respectively (Note: in the formula, the π constant comes with a + sign if $\alpha_{i \to j} \in (-\pi, 0]$ and with a - sign if $\alpha_{i \to j} \in (0, \pi]$). The coefficient $\omega \in [-1, 1]$ denotes the ahead or behind preferences of pair members: $\omega > 0$ means that agent $i$ prefers having agent behind him or her, and vice versa; $\omega = 0$ indicates a preference for walking perfectly abreast.

Fig. 4(a) shows a visualization of the pairing energy $E_{pairing}$ based on Eq. (6), as a function of the $x$ and $y$-components of the intra-pair separation. Fig. 4 illustrates that the minimum of $E_{pairing}$ aligns with the preferred configuration.

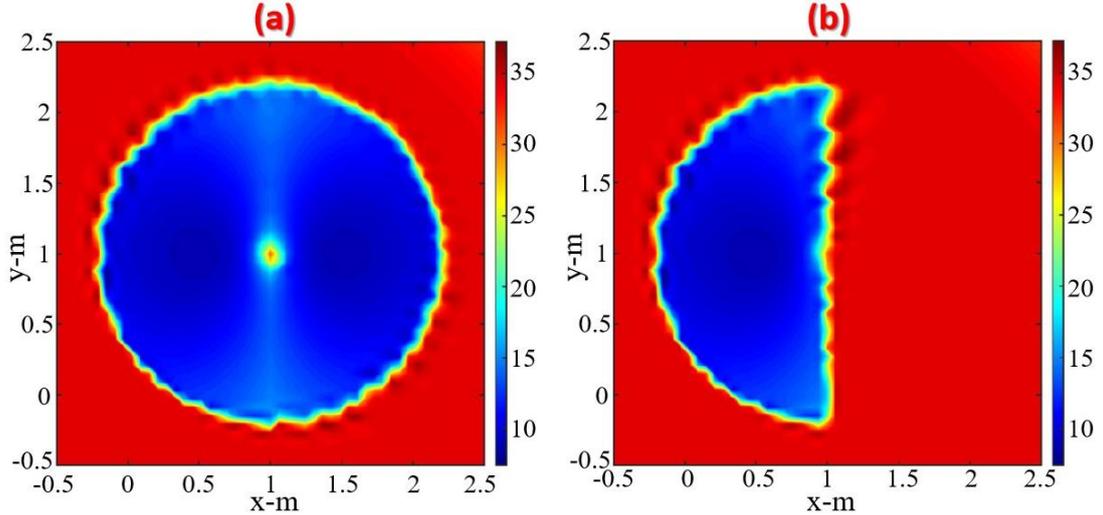

Fig. 4 Example of pairing energy $E_{pairing}$ as a function of the transverse and longitudinal components of the vector connecting the adult and child in the pair, following Eq. (6)[3]. This is done from agent $i$'s perspective, who assumes agent $j$'s velocity remains unchanged. Both panels are identical, up to the fact that panel (a) discards the constraint on the sign of the angle $\alpha_{i \to j}$. Parameters for drawing this Figure are set as: $\omega = 0$, $d_0 = 0.5$-m, $\delta t = 1$-s. Here, agent $i$ and agent $j$ are assumed to start at positions (0,0) and (1,1) respectively, with initial velocity of (0,1).

## 3. Specification and validation of the model

---

[3] For display purposes, we substitute '+∞' with 'the next largest value determined by Eq. (6)+1'.

## 3.1 Specification of the model parameters

In view of the scant quantitative data about the walking characteristics of adult-child pairs brought to light in our survey of the literature, we decided to organize a controlled experiment to determine the value of crucial pairing-related parameters for our model (see Section 2). We are mainly interested in two parameters, namely, inter-pedestrian distance and the pairs' free-walking speed $v^{pref}$, with as little disturbance from the external environment as possible.

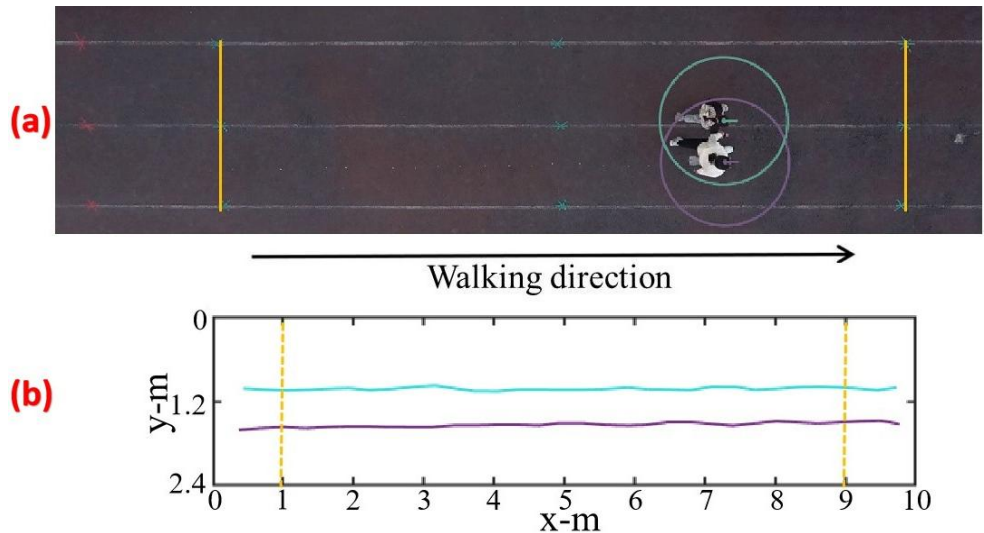

Fig. 5 (a) Snapshot of the recording of a controlled experiment involving an adult-child pair in free walk; the yellow lines materialize the starting and ending points, at x=0-m and x=10-m. (b) Trajectories extracted from the experiment with that pair; the analysis is restricted to the pseudo-stationary regime between x=1-m to x=9-m (dashed yellow lines).

The experiment was conducted at a Chinese primary school[4] on 11/10/2022 and involved seven adult-child pairs (i.e., seven adults, each holding hands with their child)[5]. Each pair was simply asked to naturally enter and exit an experimental area of rectangular shape, with length 10-m in the direction of motion and 2.4-m in the transverse direction, as sketched in Fig. 5(a); no more than one pair was present simultaneously in the rectangle, to avoid any risk of disturbance to the free-flow conditions.

---

[4] The empirical study was examined jointly by the head director of School of Transportation Science and Engineering, Beihang University, and the headmaster of Xuanhe Central Primary School, Longyan City, Fujian Province, and it received their approval.

[5] Children selected from Grade 2 were approximately 7-8 years old; the adult participants were primary school teachers, including 6 females and 1 male.

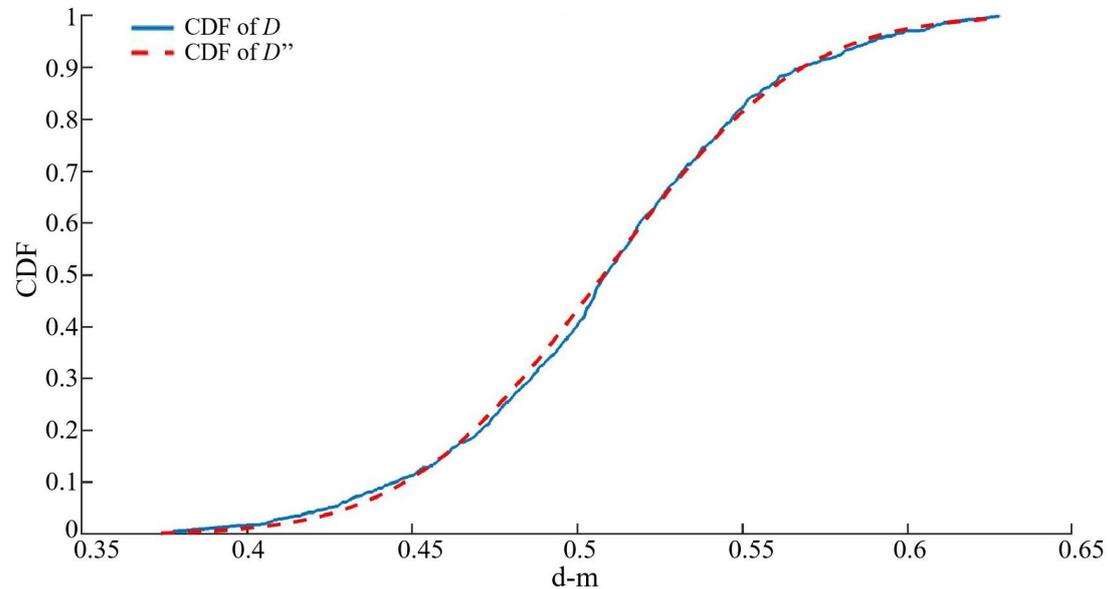

Fig. 6 Cumulative Distribution Function (CDF) of the experimental inter-pedestrian distance $D$ (solid lines), compared with the CDF of a Gaussian distribution $D''$ of mean 50.8-cm and standard deviation 4.7-cm (dashed lines).

Calibration was performed before the experiment, by positioning two participants, of the typical height of an adult and of a child, at the points marked by a cross in Fig. 5(a); the calibration, conversion from pixel coordinates to meters, and computer-aided manual tracking were carried out as in Ref. [46] (in which the experimental uncertainty on positions was less than 20-cm), using the software developed by one of us. Each participant was tracked manually every 0.4-s, on frames such as that shown in Fig. 5(a), where the light blue and purple circles added by the algorithm (delimiting the area within reach of a stretched arm) are centered on the tracked head positions of the child and adult, respectively. The extracted trajectories are drawn in Fig. 5(b), after linear interpolation to get positions every 0.1-s. The analysis below is restricted to the range x=1-m to x=9-m, to discard the transients at the beginning and at the end.

Let us first consider the Euclidean distance $D$ between adult and child. Fig. 6 shows that these distances seem to follow a Gaussian distribution of mean 50.8-cm and standard deviation 4.7-cm. Defining the normalized variable $D'=(D-0.508)/0.047$ (Unit: m), we have confirmed with a Kolmogorov-Smirnov (K-S) test [47] with a 5% significance level that no significant difference between $D'$ and a standard Gaussian distribution can be evidenced. Accordingly, we set the desired pairing distance $d_0$ entering $E_{pairing}$ in the model to 0.5-m.

Turning to the free walking speed, we estimate the mean speed over the experimental area by measuring the time needed by pairs to move between the dashed yellow lines in Fig 5(b) (x=1-m and x=9-m). We arrive at an average speed of 1.36-m/s over all pairs. The experimental conditions and limited number of participants prevent the reliable estimation of a standard deviation. Nonetheless, to preserve some variability for this crucial model parameter, we posit that the preferential speeds of agents in the model $v^{pref}$ is drawn randomly from a Gaussian distribution of mean 1.4-m/s and standard deviation 0.15-m/s, $N(1.4,0.15^2)$, and identical for both pair members.

Finally, the controlled experiment provides clear evidence of the continuous hand-holding behavior within each pair and the propensity to walk abreast. Adding to this observation the presumed innate intention of adults to oversee and protect their children while walking, the front-back coefficient $\omega$ entering $E_{pairing}$ is set to $\omega = -0.1$ for adults and $\omega = 0.1$ for children, so that adults tend to walk almost abreast with, but slightly behind, their child.

Using anatomical data, the maximal pairing distance is set to 1.2-m on the basis of the average arm length of Chinese adults (0.7-m) and children (0.5-m) [48, 49], while the body radii of adults and children are randomly drawn from $N(0.18,0.01^2)$ and $N(0.16,0.01^2)$, respectively, in line with medical statistical data [50] about the body widths of Chinese adults and children.

3.2 Numerical simulations of the model

3.2.1 Performance of the model for single agents

We open this Section with a short reminder of the capabilities of the model when it comes to reproducing the behavior of crowds made of single agents. Major successes of the model include the replication of the fundamental diagram in unidirectional and bidirectional flows, lane formation and the approximate time delay before the onset of lanes, navigation in complex geometries, and a semi-quantitative account of the enhanced chaoticity observed in bidirectional flows when some agents are 'smartphone-walking' [44].

3.2.2 Pair avoidance in dilute flow conditions

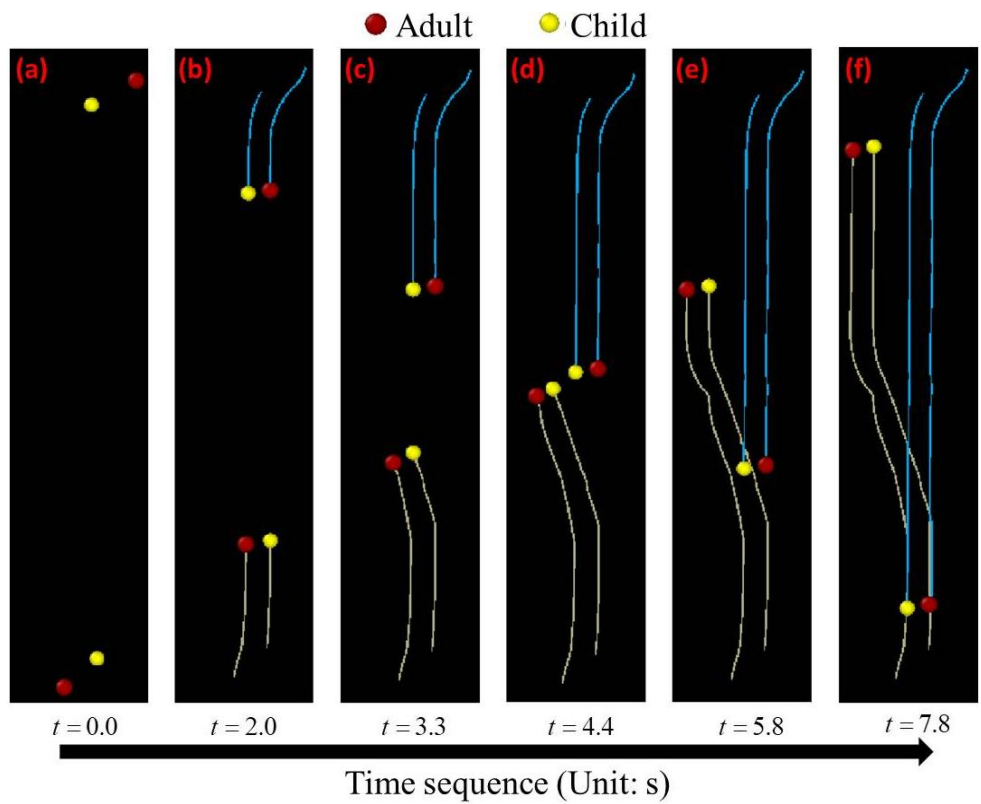

Fig. 7 Avoidance behavior of two counter-walking pairs, taken in isolation from the environment. Adults are shown in red, and children, in yellow.

Compared to simple force-based models in which collision-avoidance forces depend on relative distances, the proposed model enables agents to anticipate future potential collisions from a relatively far distance thanks to the TTC dependence of $E_{anticipation}$, and to adjust their velocities accordingly. For a more vivid demonstration of the coupled effect of $E_{anticipation}$ and $E_{pairing}$, we simulate two pairs facing each other and walking in opposite directions, within a dilute flow, and present a time sequence of their trajectories in Fig. 7. We also let adult-child pairs form spontaneously at the beginning of the simulation, $t = 0.0\text{-s}$, by initially positioning them a random distance away from each other. We observe that no sooner has the simulation started than adults and children move to form a tightly connected pair (thanks to $E_{pairing}$), which is already completed at $t = 2.0\text{-s}$. In parallel, they start moving together toward their target destination, coordinating their motion to stay abreast and united. They pass the other group around $t = 4.4\text{-s}$ but, before this, the pair with white trajectory anticipated the risk of collision (quantified by $E_{anticipation}$) and changed their course in the time window $2.0\text{-s} < t < 3.3\text{-s}$. After $t = 5.8\text{-s}$, the effect of the avoidance is no longer felt and both pairs head straight for their target. Thus, the model maintains realistic adult-child pairs even when under disturbed conditions, due to other people. We remind the reader that the non-intruding avoidance that we find (and expect) differ from the field observations of frequent intrusions made by Gregorj et al. for more general dyads of family members [43], because tight contact and hand-holding were not so widespread in their situation of study. On another touch, the effectiveness of anticipation and collision avoidance can be gauged by comparing the results of Fig. 7 (where no severe braking is observed) with the point made by Chraibi and colleagues [51] that force-based simulations of bi-directional motion may result in agents unrealistically oscillating back and forth along their direction of motion due to their conflicting moves. This artifact should be proscribed, lest deadlocks (unsolvable conflicting situations) occur in larger-scale simulations (see Fig. 12 in Ref. [38]).

3.2.3 Unidirectional flow in a free corridor

Typical sites featuring adult-child mixed flows in China are training schools. One-way corridors and T-shape junctions between corridors are found at these locations, as illustrated in the spatial layout of a Chinese training school presented in Fig. 8. We are thus interested in checking the performance of the model in generic one-way corridors without obstacles, which are idealized settings for unidirectional flow in which the microscopic and macroscopic flow properties can be quantitatively analyzed and validated.

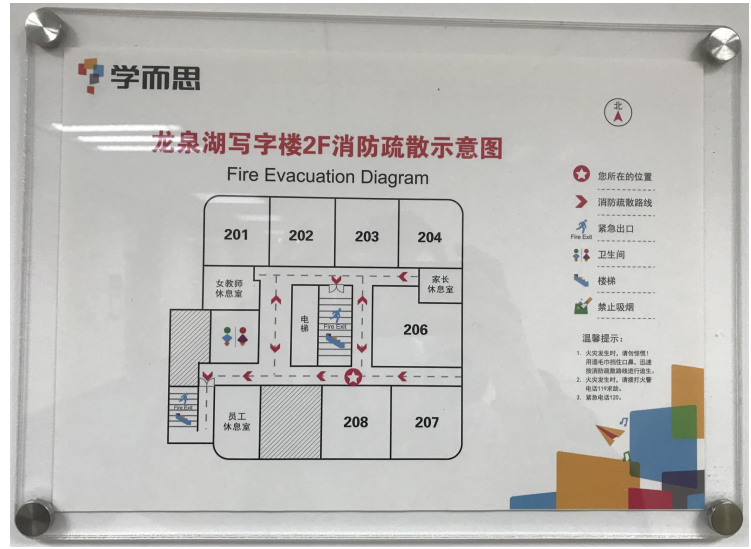

Fig. 8 Spatial layout of 2$^{nd}$ floor's corridor area of Shangdi Campus, TAL Training school, Beijing (picture taken on 1$^{st}$, January 2021).

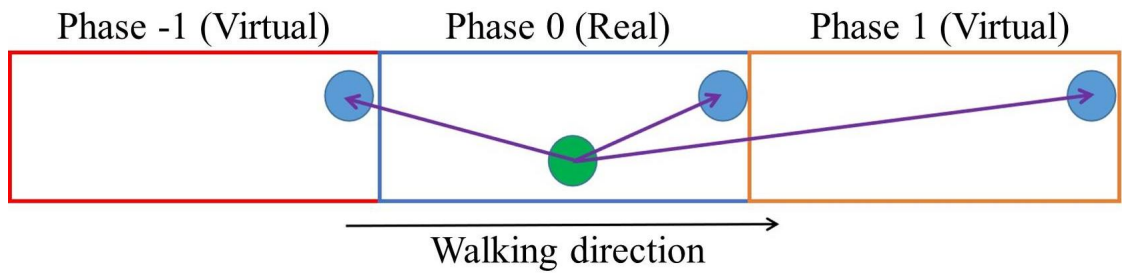

Fig. 9 Illustration of the implemented periodic boundary conditions: Phases -1 and 1 are virtual replicas of Phase 0; the green and blue agents in Phase 0 denote the one currently handled by the algorithm and one of its neighbors, respectively. The green agent interacts with all agents in Phase 0 as well as with their copies in Phases and -1.

At the collective scale, the fundamental diagram (F-D) relating walking speed to density is the most relevant indicator in practice. To evaluate it numerically, suitable continuous boundary conditions must be imposed to eliminate the variations in global density caused by pedestrians entering/exiting the experimental area. For instance, Paetzke et al. [52] and Wang et al. [53] both built a oval setting for their empirical study to maintain the global density. However, agents need to adjust the direction of velocities in a oval setup. Instead, we opt for the periodic boundary conditions depicted in Fig. 9, where Phases 1 and -1 serve as 'virtual' replicas of Phase 0. All agents except the agent currently handled by the simulation are copied in all virtual replicas and the handled agent will determine its desired velocity considering them all. After each loop, if an agent's position lies outside Phase 0, it will be translated to the equivalent position in Phase 0. Therefore, there is no need to add or remove agents during the simulation.

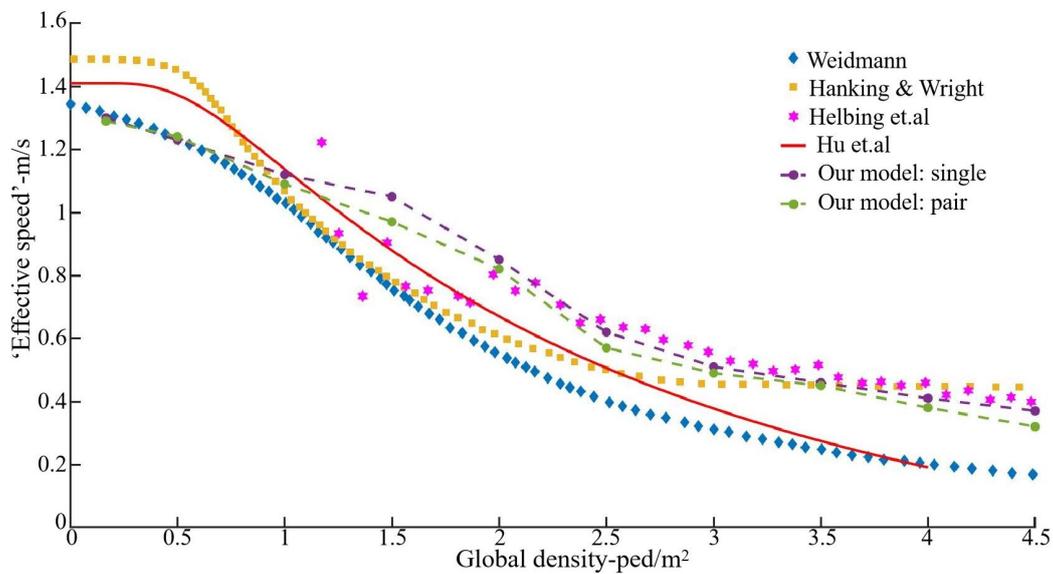

Fig. 10 Fundamental relations of uni-directional pedestrian flow of the current and some previous studies.

We simulate the unidirectional flow of single agents, on the one-hand, and adult-child pairs, on the other hand, at different densities in these conditions. The mean 'effective speed' is computed by averaging the longitudinal component of the velocity of each agent over all time steps, and then over all agents. In Fig. 10, the simulation output is compared to the empirical F-D of crowds of single adults collected at the University of Wuppertal [54] and crowds of ordinary pairs of adults collected by Hu et al. [55], noting that no empirical F-D is available for mixed pairs. There is only little difference between the F-D for single adults and for pairs. We find a reasonably good agreement between the simulated F-D and the empirical ones, in dilute flows but also at high density. The slight difference between the situations with singles and with pairs is also in line with the empirical data, as well as with the findings of Vizzari et al. [56] (for adult pedestrians).

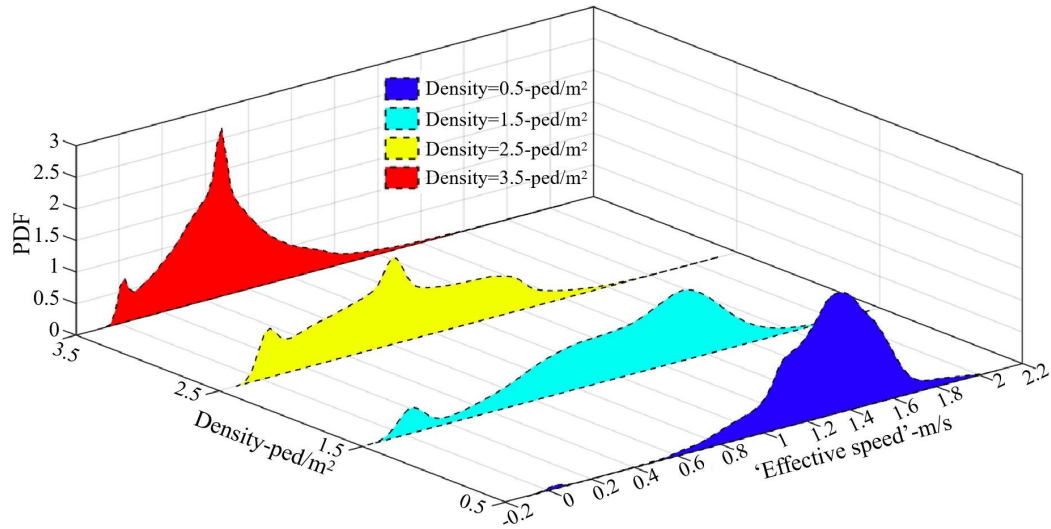

Fig. 11 PDF of the effective speeds of adult-child pairs in a simulated unidirectional flow, for four different global densities.

Beyond the coarse-grained, aggregate picture provided by the F-D, more microscopic information, the distribution of agent-based speeds is also insightful. The Probability Distribution Function (PDF) of speeds was notably used by Zanlungo et al. in their empirical study [57]. Following them, we plot the PDF of 'effective speed' at global densities 0.5-ped/m², 1.5-ped/m², 2.5-ped/m², and 3.5-ped/m² in Fig. 11. We observe that most adult-child pairs maintain their $v^{pref}$ (whose mean is 1.4-m/s) in dilute flows (0.5-ped/m²), suggesting that only few detouring or halting behaviors take place. As the global density increases, the peak around 1.4-m/s in the PDF (mostly comprised of unperturbed pedestrians) is first preserved (at density 1.5-ped/m²), while the lower tail of the PDF thickens. A secondary peak emerges close to 0-m/s, corresponding to pairs halted or moving transversely. Further increasing the density, the peak is shifted to considerably lower effective speeds, but there is still a significant fraction of people moving around 1.4-m/s. Finally, as the density reaches 3.5-ped/m², a pronounced peak around 0.4-m/s gets visible and the high-speed region of the PDF gets strongly depressed: virtually all pairs undergo detours or slowing downs in the congested corridor; many pairs appear to halt, suggesting a 'stop and go' motion.

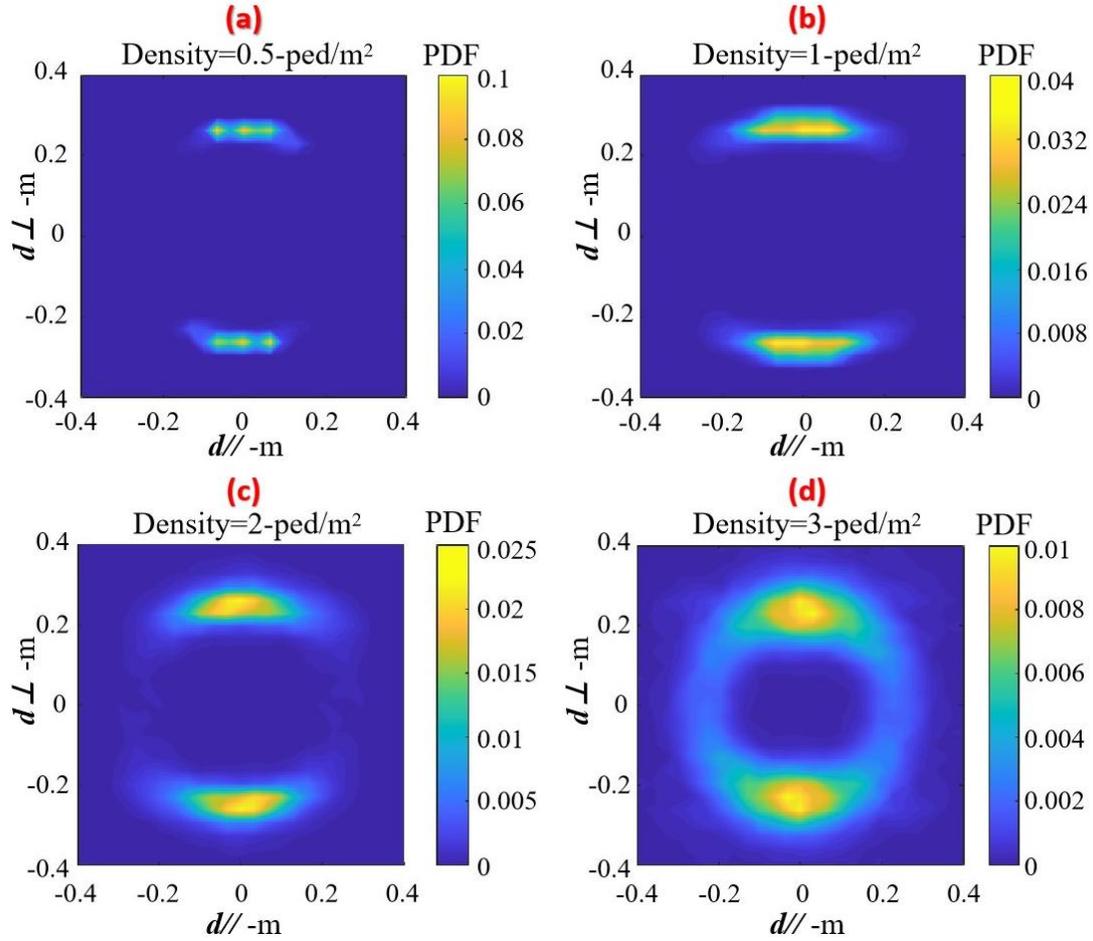

Fig. 12 The 2-D PDF for uni-directional adult-child mixed flows under various global densities.

Turning to the microscopic configuration of adult-child pairs, we choose not to consider only the inter-pedestrian distance studied by Zanlungo et al. [57], but the 2-D relative positions of pair members examined by Gorrini et al. [58] (note that both Zanlungo et al. [57] and Gorrini et al. [58] focused on homogeneous dyads), which employed a fixed coordinate system in the lab frame. To better probe the microscopic configuration within each pair, regardless of their destination, we instead resort to the pair's co-moving frame (whose origin lies at the middle point between pair members), with axes oriented relative to the instantaneous mean velocity in the pair (i.e., the average of the adult's velocity and the child's velocity). Relative positions are then decomposed into a component $d_{//}$ along the walking direction set by the mean velocity and a transverse component $d_{\perp}$; this re-orientation of the frame has the advantage that the axes are directly meaningful for pair members.

Collecting the relative positions over all time steps, we obtain the 2-D PDF shown in Fig. 12, for distinct densities: 0.5-ped/m$^2$, 1-ped/m$^2$, 2-ped/m$^2$, and 3-ped/m$^2$.

The observed PDF are qualitatively consistent with those observed empirically for dyads. The peak probabilities located at *d//*=0 indicate that pair members typically walk more or less abreast. More precisely, while the central symmetry is obtained by construction, the observed symmetries by reflection with respect to the horizontal and vertical axes result from the left-right indifference of pair members: even though the pairing energy is asymmetric between adult and child, the fact that the child can indifferently be to the right or to the left of the adult restores these symmetries. Inspecting the PDF at different densities in greater detail, we see that the pair members' intention to walk more or less abreast, which is satisfied at low density (0.5-ped/m$^2$), is more and more constrained as the density rises. Owing to the squeezes and contact pressure generated by the rest of the crowd, each pair faces more difficulties in maintaining their desired pairing format and, some may end up in a front-back configuration, notably at density 3 ped/m$^2$ (see Fig. 12(d)). Meanwhile, the distance between pair members (constrained as it is by hand-holding) does not substantially wane as the density changes, even though smaller separations are more frequently found at higher densities. The absence of points at large separations indicates that the hand-holding pairs are not separated by the (dense) flow. Overall, the microscopic information about the pairing configuration shown in Fig. 12 support the idea that the 'tight but flexible' nature of adult-child pairing is well captured by the simulations in a range of densities.

## 4. Results: Corridor-design advice and crowd-guidance strategy

Having validated several aspects of our continuous agent-based model, we now aim for practical applications in the context of egresses at training schools. More precisely, after our previous works about the interior of classrooms [6] and staircases [24], and the simulation of corridor flows in the previous section, we will put the focus on the T-shape junctions between corridors, ubiquitous in training schools (see

Fig. 8) and generating conflicting moves due to the confluence of two crowds at the junction. Our goal is to put forward guidance to practitioners about both the design of the corridors and possible strategies to guide the crowd.

Over the past years, there have been a number of endeavors to enhance the performance of emergency evacuations in primary schools, especially in the after-class period (e.g., Refs [10, 28]). But these endeavors have focused on homogeneous crowds of children, whereas, in the after-class period, the crowd often consists of adult-child pairs, which raises the level of complexity. At this stage, we must clearly admit that flows in emergency (evacuation) and non-emergency (egress) settings may substantially differ, with a different psychological status in the crowd. That being said, we may expect strategies and infrastructures that contribute to enhancing the egress performance in the after-class period to be also beneficial for evacuations.

4.1 Effect of the corridor width

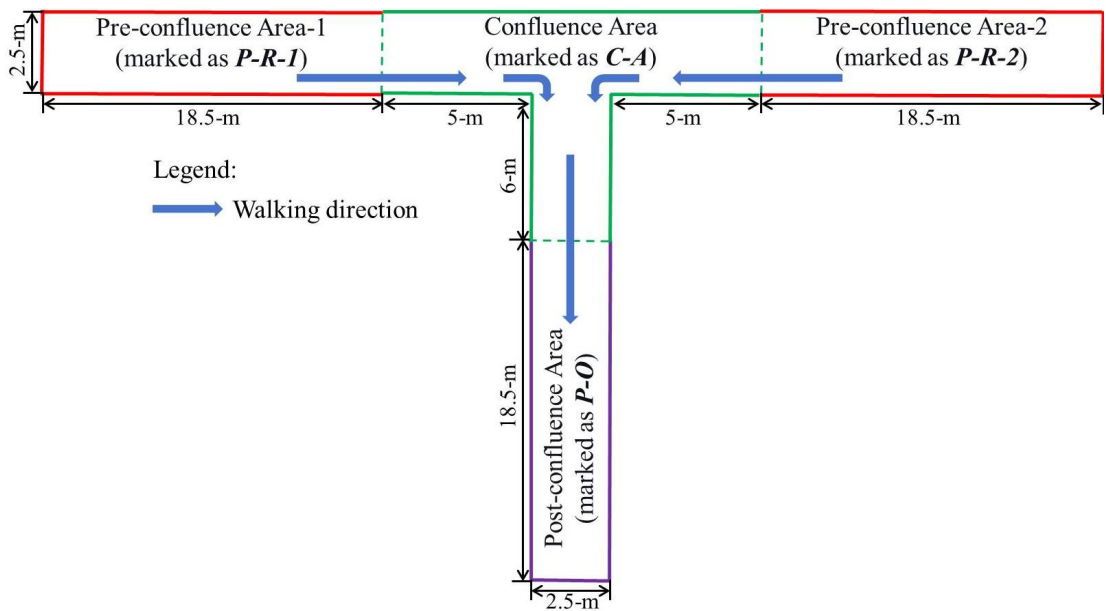

Fig. 13 Spatial layout of the simulated T-shape corridor.

Corridors in Chinese schools should comply with the Chinese national standard [59]. Accordingly, we first simulate a T-shaped junction with just the required width (i.e., 2.5-m), according to Ref. [59], for the corridors. The whole T-shape area for simulation is divided into three functional areas: two pre-confluence areas (**P-R-1** and **P-R-2**), a confluence area (**C-A**) and a post-confluence area (**P-O**), the dimensions of which are detailed in Fig. 13. At the beginning of the simulation, pairs are uniformly distributed in **P-R-1** and **P-R-2**. After their onset and gradual stabilization, the two resulting flows will converge and merge in **C-A**, and leave through **P-O**. The mixed crowd's flow at the confluence is the primary interest of this section and will be analyzed in terms of spatial occupation of the premises and inter-pair conflicts.

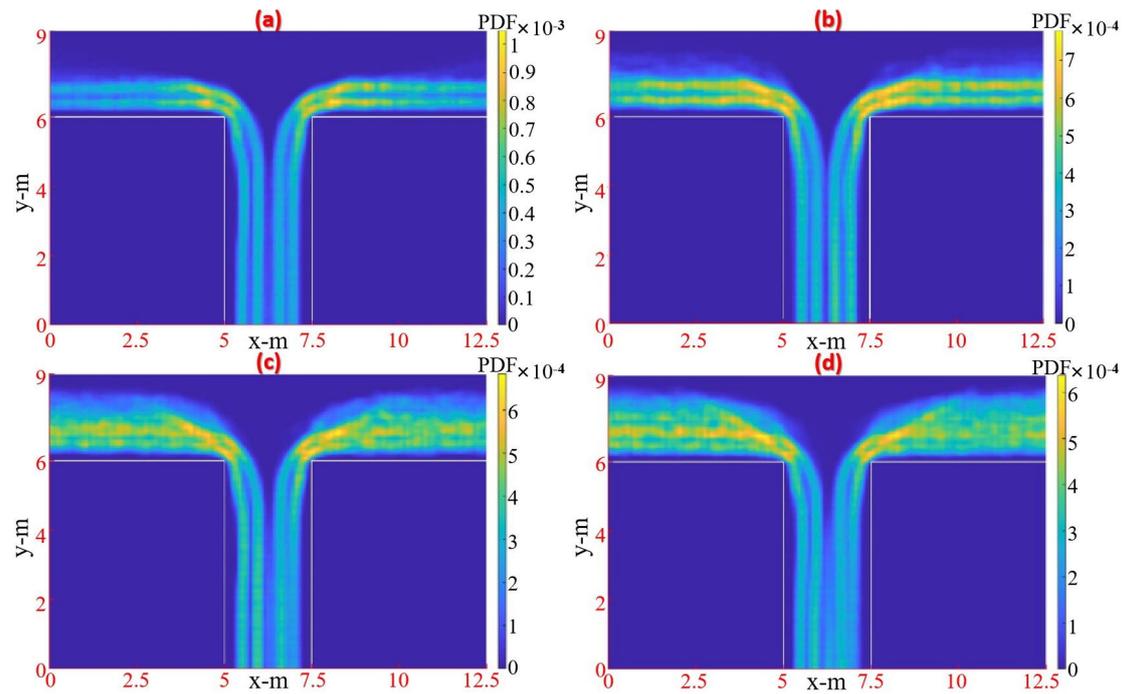

Fig. 14 Heat map of the simulated spatial occupation by adult-child pairs, for different global densities: (a) 0.65-ped/m$^2$, (b) 1.08-ped/m$^2$, (c) 1.95-ped/m$^2$, (d) 3.02-ped/m$^2$.

The average occupation of a given point in space is quantified by the relative duration during which it was covered by a disk representing an agent. To explore the effect of density on pairs' movement, cases with various initial global densities per direction: 0.65-ped/m$^2$, 1.08-ped/m$^2$, 1.95-ped/m$^2$, 3.02-ped/m$^2$, i.e., (number of agents in **P-R-1**)÷(area of **P-R-1**) or (number of agents in **P-R-2**)÷(area of **P-R-2**), are simulated (see Fig. 14). Note that the maximal density under study, 3.02-ped/m$^2$ remains substantially below tight-packing (even if there were only adults) and even below the regime of strong contacts (typically around 6-ped/m$^2$), at which our assumptions of circularly shaped agents would really reach their limitations. Taking Fig. 14(a) as an example, both **P-R-1** and **P-R-2** are assigned with 0.65-ped/m² as the initial density to make sure the pair's distribution before confluence is balanced. From Fig. 14, one can see that most pairs can choose relatively optimal paths toward the destination in low-density cases (Figs. 14(a) and (b)); with the rise of global density, the space-occupying area also inflates gradually (Figs. 14(c) and (d)), which denotes that some pairs need to make detours. Meanwhile, Figs. 14(c) and (d) show that pairs take into account future collision with opposite pairs (owing to $E_{anticipation}$) and consequently stay fairly close to the wall on their side, even after confluence. This leads to distinctive features i) a lane formation phenomenon; ii) a cone-shaped empty area around the confluence central point. Interestingly, such an area was also reported by Zhang et al. [60]. Moreover, the wall-agent inter-distance after confluence in Fig. 14 is similar to the empirical results of Zhang et al. [60].

To analyze the friction ('conflicts') in the flow, we build on the private space invasion variable $\varepsilon$ defined in the model (see above), which reflects the agents' will to preserve a private space around themselves. For clearer results, the spatial extent for $\varepsilon^*$ of the private space (entering Eq. (3)) was doubled for the representation of Fig. 15, compared to the actual simulations[6]. Importantly, the overlapping space between agents in the same pair is not considered. The spatial distribution and strength of overlaps between the 'extended private spaces' of pair members are presented in Fig. 15, pointing to the magnitude of inter-pair 'conflicts'. Clearly, conflicts do not distribute randomly in space; on the contrary, the distribution pattern in Figs. 15(a)-(d) roughly follows the occupation of space revealed by Figs. 14(a)-(d): not surprisingly, denser regimes and denser areas entail more conflicts. Because of $E_{anticipation}$, the central part of the confluence area is seldom chosen by pairs, hence rare conflicts there. Additionally, one can see that conflict happens more significantly within space corresponding to the starting phase of the confluence process rather than the ending phase; such collective phenomenon demonstrates that in stabilized unidirectional flow, friction is reduced, whereas before confluence the moves in preparation of the turn generate conflicts.

---

[6] $\varepsilon$ value is double only for drawing Fig. 15, while it remains unchanged for the simulation.

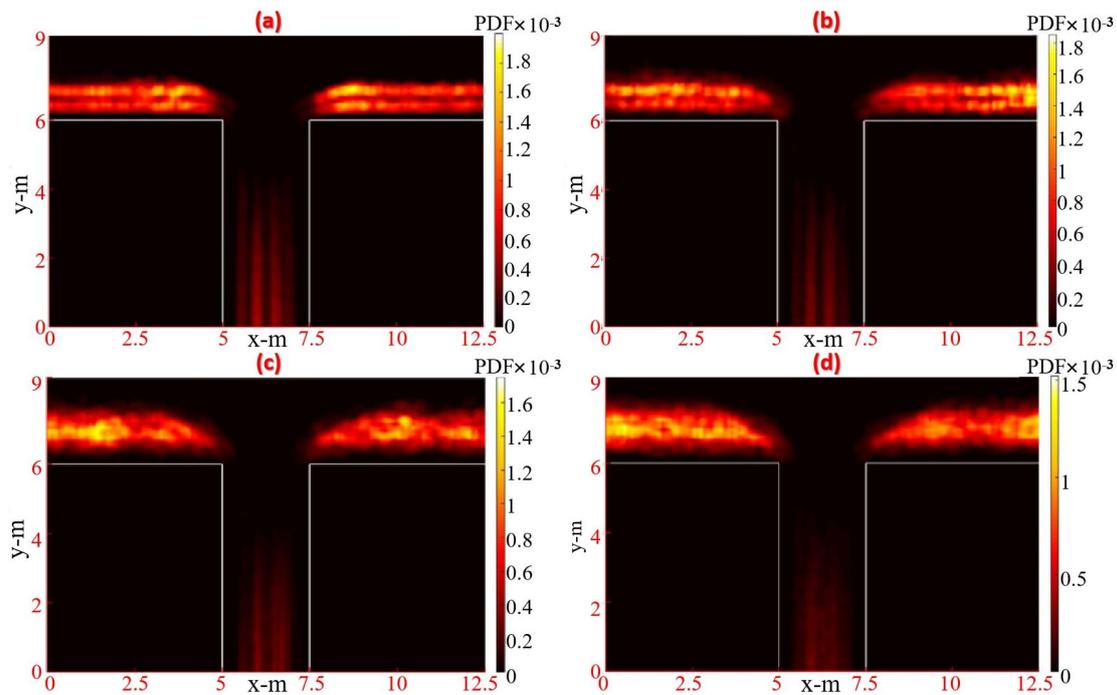

Fig. 15 The 2-D PDF of inter-pair spatial conflict in the T-shape corridor, where (a)-(d) correspond to Figs. 14(a)-(d), respectively.

Although the Chinese national standard regulates the corridor-width design, a large number of training schools still fail to implement the standard, owing to the limitation and improper arrangement of space (as disclosed by Chinese government reports [5, 61]). Tragedies occurred because of poor design and management of space, notably [62], hence the importance of studying the impact of corridor width on egress/evacuation performance. In an empirical study, Ren et al. revealed that different widths of a straight corridor entailed discrepancies in the flow of crowds of elderly adults [63], but the more complex case of adult-child mixed flows deserves to be investigated separately. To this end, we narrow **P-O**'s width to 1.5-m in our simulations while keeping the rest of the layout unchanged. The heat maps of space occupation and inter-pair spatial conflict are presented in Figs. 16 and 17. Unlike Fig. 14, the

*double* lane structure is no longer observed after confluence in this narrower corridor: pairs coming from ***P-R-1*** and ***P-R-2*** compete at confluence and the pairs that fail to occupy space first must lower their speed (or come to a halt) while waiting for the next chance. Such competition initially causes congestion around the corner area, and pairs further accumulate within the straight-walking areas (i.e., ***P-R-1*** and ***P-R-2***) over time. Although the cone-shaped empty area still can be partly observed at low density (see Figs. 16 (a) and (b)), with rising global density, more and more pairs accumulate in the central-confluence area due to congestion (see Figs. 16 (c) and (d)), causing both the disappearance of this empty area, as well as an upwards shift of the central convergence point (the red dot in Figs. 16(a)-(d)).

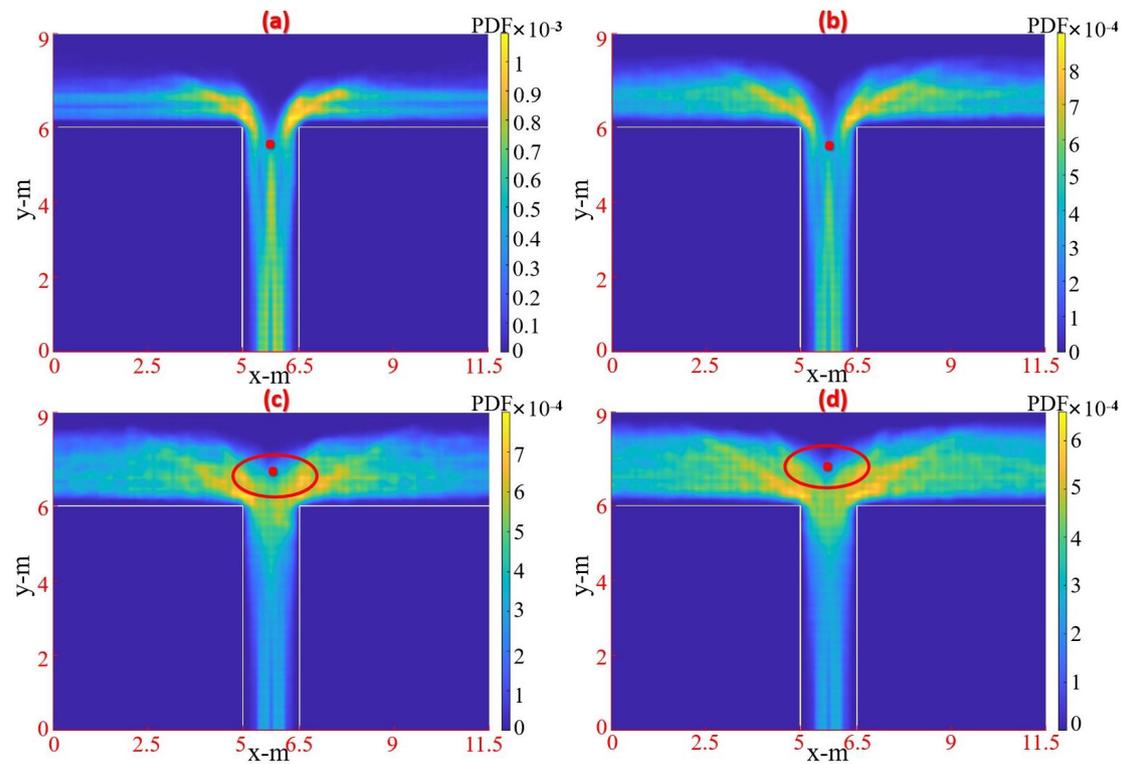

Fig. 16 Heat maps of space occupation with the ***P-O*** corridor width narrowed down to 1.5-m, where (a)-(d) denote cases with initial global density equaling 0.65-ped/m$^2$, 1.08-ped/m$^2$, 1.95-ped/m$^2$, 3.02-ped/m$^2$, respectively.

The spatial distribution of inter-pair 'conflict' (see Fig. 17) also reveals the direct and indirect effect induced by a narrower ***P-O*** corridor. For one thing, although the spatial conflict still has a similar distribution pattern with the space occupancy, conflicts occurring around corners and within the central area of *C-A* in Fig. 17 is more evident than the counterpart in Fig. 15. For another, conflicts will occur more intensely and frequently in the pre-confluence corridor (i.e., *P-R-1* and *P-R-2*) owing to the upstream propagation of the jam at confluence. It is generally known that crowd congestion and inter-agent conflict belong to the main reasons which may affect the efficiency and safety of crowd movement in critical conditions; Figs. 16 and 17 quantitatively illustrate that a narrower width design of the T-shape corridor negatively influences the mixed crowd's movement. The situation may worsen and turn more chaotic when adults are absent. Consequently, the requirement of minimal corridor widths significantly wider than 1.5-m by the Chinese national standard [59] appears a sensible measure for safe and efficient egresses with the inflows typically encountered at training schools.

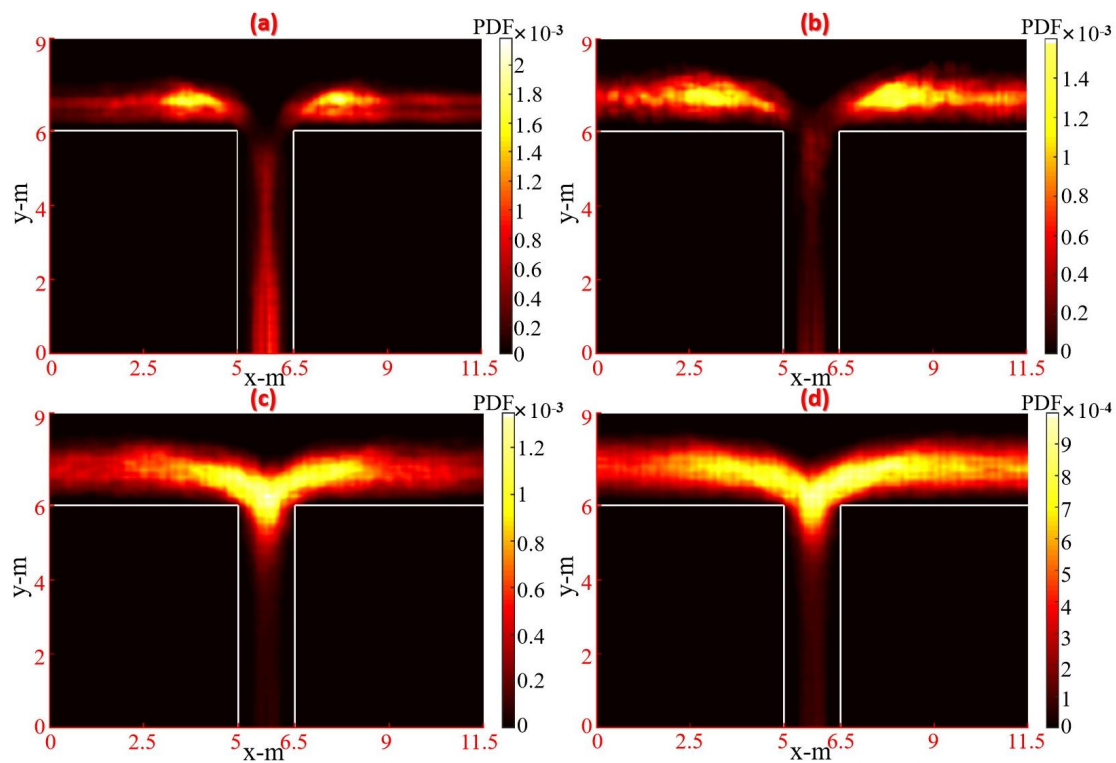

Fig. 17 Heat maps of "conflicts' (friction) between pairs in the T-shape junction with a corridor narrowed down to 1.5 m, where (a)-(d) correspond to the same global densities as Figs. 16(a)-(d), respectively.

4.2 Effect of pairs' hand-holding behavior

Tab. 1 Average inter-pedestrian distance (*AID*, tentative safety indicator) and average travel-time (*ATT*, efficiency indicator) for 'solid' and 'loose' crowds at various initial global densities.

|  |  | Global density | | | |
| --- | --- | --- | --- | --- | --- |
|  |  | 0.65-ped/m$^2$ | 1.08-ped/m$^2$ | 1.95-ped/m$^2$ | 3.02-ped/m$^2$ |
| *AID* | 'Solid' bonds | 48.2-cm | 48.4-cm | 48.1-cm | 47.9-cm |
|  | 'Loose' bonds | 48.6-cm | 50.4-cm | 66.1-cm | 82.2-cm |
| *ATT* | 'Solid' bonds | 9.69-s | 11.35-s | 13.85-s | 14.64-s |
|  | 'Loose' bonds | 9.72-s | 11.39-s | 14.19-s | 15.34-s |

Daily-life observations comforted by the empirical study in this paper have led to the implementation of hand-holding between child and adult, via the pseudo-energy $E_{pairing}$, which captures this behavior in a rather realistic way (Fig. 12). Nevertheless, social groups consisting of ordinary adults usually tend not to hold hands while walking. But one can also consider situations in which pairs do not hold hands, as in other social groups. We will refer to adult-child pairs systematically holding hands as 'solid' pairs, in contrast with the 'loose' pairs intent on holding hands, but not necessarily doing. How do the collective dynamics then differ between crowds of solid pairs and crowds of loose pairs? Here, the 2.5-m-wide T-shape confluence corridor of Fig. 13 is restored. By recording the entrance and exit time of each pair in and out of **C-A**, we obtain their travel time through **C-A**, which we then average over the simulated pairs into an average-travel-time (***ATT*** in short). In addition to this efficiency indicator, a safety indicator is needed. Drawing inspiration from a UNICEF article [64], which emphasizes the importance of physical safety guidelines delivered from parents to their children during emergencies, we regard the separation between adult and child as a critical hazard under emergency conditions and set the average-intra-pair-distance (***AID*** in short) as a tentative safety indicator for our simulations, even though they are conducted in non-emergency settings. ***AID*** is readily obtained by collecting the inter-pedestrian distance in each pair at every update time and averaging these distances. The values of the *AID* and *ATT* indicators for 'solid' vs. 'loose' crowds at diverse global densities are shown in Tab. 1.

Interestingly, Tab. 1 reveals that, at low enough densities (below 1-ped/m$^2$) 'solid' and 'loose' crowds share similar values for the efficiency and safety indicators, as expected. As the crowd gets denser, the *ATT* travel time rises, increasing by 50% between global density 0.65-ped/m$^2$ and 3-ped/m$^2$ for both types of crowds. The *ATT* of loosely bonded crowds increases somewhat more sharply, reflecting hindrances in the flow due to split pairs, but overall the difference is relatively small (less than 1-s). In contrast, the *AID* intra-pair distance hardly varies for solid pairs, apart from a very minor decrease due to the density, who are forced to keep holding hands (thus, adults are still able to guide/assist their children), whereas *AID* dramatically increases for loose crowds, by more than 30-cm between 1-ped/m$^2$ and 3-ped/m$^2$, confirming that in this situation pairs in dense crowds may end up split, potentially imperiling children. Note that involuntary splitting of groups who are intent on staying together has been reported in critically dense crowds, for instance during the 2010 Love Parade crowd crush [65]. In summary, for the adult-child mixed crowd movement in the T-shape confluence corridor, both the safety and (to a much more minor extent) the efficiency indicators are better for 'solid' pairs: our simulations bolster the idea that encouraging adults to tightly hold their children's hand can help to achieve safer and (marginally) quicker egresses. In the conclusion below, we shall put in perspective these noteworthy numerical findings in the context of public safety in real life.

## 5. Summary and discussion

Prompted by recurrent stampedes in crowded school-like environments during the after-class period (especially in developing countries), this paper is dedicated to modeling the pedestrian flow of adult-child pairs. These mixed flows are frequently observed, notably at Chinese training schools, and display distinctive features, such as hand-holding and heterogeneous body sizes, which impact the dynamics, but existing models fail to satisfactorily account for these features. Thus, in light of the interest of simulations (especially whenever ethical concerns would hinder realistic experiments), the primary objective of this paper was to fill this modeling gap, with a practical focus on enhancing safety and efficiency in facilities where adult-child pairs naturally assemble, more specifically, on (unobstructed) training schools corridors. This endeavor required a substantial extension of a recently proposed and validated pseudo-energy-based pedestrian model for homogeneous crowds. The incorporation of a distance and angle-dependent pairing energy $E_{pairing}$, specifically tailored to depict the 'tight but flexible' nature of adult-child pairs, led to a successful replication of empirical data on single pairs in free flow. Qualitatively realistic features of the individual and collective dynamics (side-by-side formation, relative positions of adult and child at different densities, pair avoidance behavior, motion at a corridor angle) were also reproduced, along with quantitative ones (namely, the fundamental diagram). Such comprehensive validation positions the model as a flexible tool test and assess corridor-design advice and crowd-guidance strategy to enhance egress safety and efficiency, and also serve as a reference situation to evaluate emergency evacuation performance. In this study, by simulating pair movements in T-shaped corridors, where two mixed flows merge, forming a post-confluence uni-flow, it is observed that maintaining the width of the post-confluence area at least equal to the pre-confluence area is key for egress efficiency, by easing congestion and reducing inter-agent collisions. In terms of crowd-guidance strategies within this T-shaped confluent environment, our findings suggest that encouraging adults to tightly hold their children's hands (i.e., to form 'solid' bonds) can help improve the safety and efficiency of the pedestrian flow, by limiting the risks of splitting pairs.

Therefore, our paper partially responds to the challenge of modeling heterogeneous pedestrian flow, specifically adult-child mixed flows, in spaces with minimal infrastructure-imposed regulations. The relevance of these models integrating mixed flows for public safety is evident in cases where experiments or drills in realistic emergency conditions would be unethical, which is particularly the case when children are involved; the improvements targeted in this paper are all the more necessary as models that fail to account for these characteristics are routinely used for practical applications, to design and dimension new public spaces.

On the other hand, there is no doubt that, for the promotion of official guidelines and recommendations at the national or international scale, numerical simulations, even performed with an improved model, are not sufficient. They are useful for pointing to management interventions that may efficiently address current regulatory deficiencies, but extensive experimental validation will be needed to establish the effectiveness of the most promising interventions before applying them widely.

Along these lines, we would like to mention certain limitations and potential advancements of this work. Firstly, our study is limited to controlled experiments at a very small scale for the extraction of features of adult-child pairs; ideally, real-life videotaping of the after-class period in Chinese training schools' corridors with accurate calibration (or more extensive controlled experiments) would yield more comprehensive insights. Secondly, in modeling the adult-child mixed flow, we presuppose a consistent one-to-one adult-to-child pairing mode, neglecting less frequent cases, such as, alone-walking individuals or different group compositions, which, though rare, do occur in training schools. Finally, we recall that the strategies suggested by our model will deserve empirical validation prior to widespread implementation, to ensure their effectiveness and uncover any unforeseen effects. Pursuing these avenues would be a natural progression from this initial research.

**Acknowledgement**

(1) Chuan-Zhi (Thomas) Xie and Tie-Qiao Tang acknowledge the following funding information: i) National Natural Science Foundation of China (72171006, 72231001, 72288101); ii) China Scholarship Council (202006020212).

(2) Alexandre Nicolas acknowledges funding in the frame of French-German research project MADRAS funded in France by the Agence Nationale de la Recherche (grant number ANR-20-CE92-0033), and in Germany by the Deutsche Forschungsgemeinschaft (grant number 446168800).

(3) The authors wish to thank Dr. Iñaki Echeverría-Huarte (Postdoctoral Researcher at Universidade de Lisboa, Lisbon), Mr. Oscar Dufour (Ph.D. Candidate at CNRS & Univ. Lyon 1, Lyon) for their modeling help and Mr. Botao Zhang (Ph.D. Candidate at City University of Hong Kong, Hong Kong) for his assistance of video-data collecting.

**Authorship contribution statement**

Chuan-Zhi Xie: Conceptualization, Methodology, Software, Investigation, Experiment, Data analysis, Writing-original draft. Tie-Qiao Tang: Conceptualization, Experiment, Supervision. Alexandre Nicolas: Conceptualization, Methodology, Software, Data analysis, Writing-polishing, Supervision.